\documentclass[%
 reprint,
 amsmath,amssymb,
 aps,
onecolumn,
]{revtex4-1}

\usepackage{graphicx}
\usepackage{dcolumn}
\usepackage{bm}

\newcommand{\bea}{\begin{eqnarray}}	
\newcommand{\eea}{\end{eqnarray}}
\newcommand{\be}{\begin{equation}}	
\newcommand{\ee}{\end{equation}}
\newcommand{\beq}{\begin{equation}}	
\newcommand{\eeq}{\end{equation}}

\newcommand{\Z}{{\mathbb Z}}
\newcommand{\C}{{\mathbb C}}

\newcommand{\vev}[1]{\left\langle{#1}\right\rangle}

\def\R{\relax\ifmmode {\mathbb R}  \else${\mathbb R}$\fi}
\def\C{\relax\ifmmode {\mathbb C}  \else${\mathbb C}$\fi}
\def\Z{\relax\ifmmode {\mathbb Z}  \else${\mathbb Z}$\fi}
\def\N{\relax\ifmmode {\mathbb N}  \else${\mathbb N}$\fi}
\def\I{\relax\ifmmode {\mathbb I}  \else${\mathbb I}$\fi}

\begin{document}


\title{ Infrared massive gluon propagator from a BRST-invariant Gribov horizon in a family of covariant gauges}

\author{Bruno W. Mintz}
\author{Leticia F. Palhares}%
\author{Giovani Peruzzo}%
\author{Silvio P. Sorella}
\affiliation{UERJ $-$ Universidade do Estado do Rio de Janeiro,\\
Departamento de F\'isica Te\'orica, Rua S\~ao Francisco Xavier 524,\\
20550-013, Maracan\~a, Rio de Janeiro, Brasil}



\date{\today}

\begin{abstract}

A distinctive feature  observed in lattice simulations  of confining nonabelian 
gauge theories, such as Quantum Chromodynamics, is the presence of a
dynamical mass for the gauge field in the low energy regime of the 
theory. In the Gribov-Zwanziger framework in the Landau gauge, such
mass is a consequence 
of the generation of the dimension two condensates $\vev{A_\mu^aA_\mu^a}$ and 
$\vev{\bar\varphi_\mu^{ab}\varphi_\mu^{ab}-\bar\omega_\mu^{ab}\omega_\mu^{ab}}$, 
where $A$ is the gluon field and the fields $\bar\varphi$, $\varphi$, 
$\bar\omega$, and $\omega$ are Zwanziger's auxiliary fields. 
In this work, we show that, in the recently developed BRST-invariant 
version of the Refined Gribov-Zwanziger theory, these condensates can 
be introduced in a BRST-invariant way for a family of $R_\xi$ 
gauges.  Their values are explicitly computed to first order and turn out to be independent of the gauge 
parameters contained in the gauge-fixing condition, as expected from the BRST invariance of the formulation.  This fact
supports the possibility of a gauge-parameter independent  nonzero infrared gluon mass, 
whose value is the same as the one in the Landau gauge.



\end{abstract}

\maketitle


\section{\label{sec:introduction}Introduction}

Since more than fifty years, the standard way to perform the quantization 
of a gauge field theory in the continuum is that of the Faddeev-Popov 
procedure \cite{Faddeev:1967fc}. One of the main features of this approach 
is that quantization is effectively performed only if some given gauge condition
is imposed. The most popular choices of gauge conditions in the study of 
the strong interaction include not only covariant and color-symmetric 
gauges, such as the Landau, the Linear Covariant, and the $R_\xi$ 
gauges \cite{Peskin:1995ev,Weinberg:1995mt}, but also noncovariant 
ones, such as the Coulomb gauge
\cite{Zwanziger:2004np,Guimaraes:2015bra,Campagnari:2016wlt} . 
Another category of gauges includes those in which 
color symmetry is not explicitly enforced, such as the Maximal Abelian 
Gauge \cite{tHooft:1981bkw,kronfeld:1987ri,Kronfeld:1987vd}. For an interesting discussion about several issues that may appear in the usage of different gauges, see
\cite{Gaigg:1990si}.

Each of these gauge choices can, at least in principle, 
facilitate the understanding of some given physical aspect of the theory. 
In any case, one expects that the actual physical observables do not 
depend on the choice of the gauge condition, so that every 
gauge choice should lead to the same physical results.  Although this 
has been shown to be true in perturbative calculations, the proof of gauge equivalence 
is not as straightforward at the nonperturbative level. Finally, 
although it is possible to successfully formulate a gauge invariant 
version of Yang-Mills theory on a discretized spacetime in the 
nonperturbative regime \cite{Wilson:1974sk}, continuum approaches still
rely on the aforementioned gauge-fixing procedure. 

As discussed by Gribov in his seminal work \cite{Gribov:1977wm}, 
the Faddeev-Popov procedure 
is consistent only at the perturbative level. For definiteness, let us 
consider the Landau gauge \footnote{An analogous argument can also be 
cast in the other mentioned gauges.}. As one considers physical processes at lower 
and lower energies, the strong coupling becomes larger and the gauge field configurations may depart from the vicinity of $A=0$. As 
a result, the Faddeev-Popov operator  $[{\cal M}(A)]^{ab}$ in the Landau gauge, $\partial_\mu A^a_\mu=0$,   
\begin{eqnarray}
 [{\cal M}(A)]^{ab} &=& -\partial_\mu\,D_\mu^{ab}\nonumber\\
               &=& -\delta^{ab}\partial^2 + gf^{abc}A_\mu^c\partial_\mu,  \label{fpop}
\end{eqnarray}
whose determinant appears in the path integral formulation, acquires zero modes, i.e., eigenfunctions with vanishing eigenvalues, rendering thus the quantization procedure ill-defined.
The presence of these zero modes indicates that some gauge copies were not 
removed by the Faddeev-Popov procedure. In other words, the gauge-fixing 
procedure has not fully eliminated the multiple counting of physical states in the path integral due to gauge invariance.   This is called the Gribov problem.

A solution to the Gribov problem would be to evaluate the path integral in such a way that only one representative for each 
gauge orbit is accounted for. This subset of the gauge field configuration 
space is the {\it fundamental modular region}, whose practical implementation 
has not yet been achieved. In order to take into account the existence of the gauge copies, 
a partial solution was proposed by Gribov himself in the Landau gauge, amounting to constrain the functional integration over the gauge fields to a 
subset of all field configurations  so that the Faddeev-Popov operator ${\cal M}$, eq.\eqref{fpop},  has only positive 
eigenvalues. This subset is called the {\it Gribov region} and its 
boundary is called the {\it Gribov horizon}.

A practical way to implement the restriction of the path integral to the 
Gribov region has been proposed by Zwanziger in \cite{Zwanziger:1989mf}. 
The  overall procedure can be seen as the use of the Lagrange multiplier method with the addition of an   extra term 
to the action to enforce the constraint of the Gribov region. This term 
is the {\it horizon function}, given by
\begin{eqnarray}\label{eq:horizon-funct-landau}
 \gamma^4H(A) = \gamma^4g^2\int\,d^dx\,d^dy\,f^{abc}A^b_\mu(x)[{\cal M}(A)^{-1}]^{ce}(x,y)f^{ade}A_\mu^d(y)
\end{eqnarray}
in the Landau gauge, where $\gamma$ is the {\it Gribov parameter},
which is not a free parameter, but rather fixed self-consistently from a gap 
equation, called the horizon condition, see \cite{Zwanziger:1989mf,Vandersickel:2012tz}.
In an initially scale-invariant theory, as is the case of 
Landau-gauge Yang-Mills, the Gribov parameter can be shown to be proportional 
to the renormalization-group invariant scale that appears as a consequence of 
conformal anomaly, leading to a nonperturbative infrared 
effective action, known as the Gribov-Zwanziger action.  Since  \eqref{eq:horizon-funct-landau} contains 
the inverse of a field-dependent  differential 
operator, the horizon function is a nonlocal functional of  the fields. 
In order to write the resulting effective field theory with the 
horizon constraint as a local field theory, it is necessary to 
introduce auxiliary bosonic fields, $\bar\varphi$ and $\varphi$, 
and fermionic fields, $\bar\omega$ and $\omega$
\cite{Zwanziger:1989mf,Vandersickel:2012tz}. With these ingredients, 
one can write down the Gribov-Zwanziger (GZ) action 
in  local form in euclidean space for the Landau gauge,  namely 
\begin{eqnarray}
 S_{GZ} = S_{FP} + S_{h}\, ,   \label{gzact}
\end{eqnarray}
where
\begin{eqnarray}\label{eq:FP-action}
 S_{FP} = \int\,d^dx\,\left[\frac14F_{\mu\nu}^aF_{\mu\nu}^a 
 + ib^a\partial_\mu A^a_\mu + \bar c^a\,\partial_\mu D_\mu^{ab}c^b\right]
\end{eqnarray}
is the Faddeev-Popov lagrangian in the Landau gauge and
\begin{eqnarray}\label{eq:horizon-action}
 S_h = \int\,d^dx\left[\bar\varphi_\mu^{ab}[{\cal M}(A)]^{ac}  \varphi_\mu^{cb} - \bar\omega_\mu^{ab}[{\cal M}(A)]^{ac}  \omega_\mu^{cb} 
 + ig\gamma^2f^{abc}A_\mu^a(\varphi^{bc}_\mu + \bar\varphi_\mu^{bc})\right]
\end{eqnarray}
is the local version of the horizon function in terms of Zwanziger's 
auxiliary fields.

An interesting feature of the gluon propagator in the GZ framework is that 
it violates reflection positivity, which is one of the Osterwalder-Schrader 
axioms of euclidean field theory \cite{Osterwalder:1973dx,Osterwalder:1974tc}.
Actually, the so-called positivity violation  has been  interpreted as
a sign of confinement \cite{Oehme:1994pv,Nishijima:1996ji,Alkofer:2000wg}.
However, the gluon GZ propagator still lacks an important feature: 
according to several results from different nonperturbative methods  -- such as
lattice numerical simulations, functional renormalization group and Dyson-Schwinger equations --, the gluon propagator should be finite 
at zero momentum and the ghost propagator should behave as $p^{-2}$
as one approaches the infrared. This is known as the decoupling solution 
\cite{Cucchieri:2007rg,Fischer:2008uz,Oliveira:2008uf,Cucchieri:2008fc,Dudal:2007cw,Dudal:2008sp,Aguilar:2008xm,Boucaud:2008ji,Cyrol:2016tym}.
The GZ  gluon  propagator, on its turn, vanishes in the infrared and the ghost propagator 
behaves as $p^{-4}$ for $p\rightarrow0$, which is called the scaling behavior 
\cite{Alkofer:2000wg}. This is an indication that the GZ
action is still plagued by some non-perturbative instability. 
In fact, it turns out that the
condensation of mass dimension $d_m=2$ operators, such as $A^2$ and 
$\vev{\bar\varphi\varphi-\bar\omega\omega}$ takes place \cite{Dudal:2008sp,Dudal:2011gd}, 
resulting in a gluon propagator which is non-vanishing at $p=0$. 
Since these condensates are nonzero,
it is possible to include them directly in the infrared action, which is 
equivalent to give mass to the gluon and to the Zwanziger auxiliary 
fields. The resulting action is called the Refined Gribov-Zwanziger 
(RGZ) theory in the Landau gauge
\begin{eqnarray}\label{eq:action-RGZ}
 S_{RGZ} = S_{FP} + S_h + \int\,d^dx\,\left[\frac{m^2}{2}A_\mu^aA_\mu^a
 + M^2(\bar\varphi_\mu^{ab}\varphi_\mu^{ab} - \bar\omega_\mu^{ab}\omega_\mu^{ab})\right].
\end{eqnarray}

In spite of the success of the RGZ approach in the Landau gauge, its 
original formulation does not allow  for an extension to other gauges
such as the Linear Covariant gauges or the $R_\xi$ gauges. The main 
reason for that is the lack of BRST invariance of both GZ and RGZ actions, eqs.\eqref{gzact} and  
(\ref{eq:action-RGZ}), respectively.  Recently, however, an equivalent formulation of the theory was developed
which possesses full BRST invariance, making the extension to 
other gauges possible \cite{Capri:2015ixa}. Indeed, some features of 
the GZ and the RGZ theories in the linear covariant gauge  have already 
been explored, for example, in \cite{Capri:2016aqq,Capri:2016gut,Capri:2017abz}.
Furthermore, as discussed in \cite{Capri:2018ijg}, the horizon function
of a large class of covariant and color invariant gauge conditions, which  
are continuously connected to the Landau gauge, can 
be expressed in a unified manner. As a consequence of this new formulation,
correlation functions of gauge invariant operators become independent 
of gauge parameters 
and are given by their values in the Landau gauge \cite{Capri:2018ijg}.

In this work, we explore the instability of the Gribov-Zwanziger 
theory that gives rise to the Refined GZ theory in a  gauge-parameter  independent 
manner.  We do this by 
considering a $R_\xi$ gauge, which can be understood 
as a generalization of the Linear Covariant gauge.
In Sec. \ref{sec:BRST-inv-GZ}, we review the BRST invariant 
local formulation of the Gribov-Zwanziger effective theory 
\cite{Capri:2016aqq} and formulate it in the $R_\xi$ gauge. Next, in 
Sec. \ref{sec:vac-energy}, we calculate the vacuum energy in the GZ 
effective theory at leading order and show that it is independent of 
the gauge parameters. Finally, in Sec. \ref{sec:condensates}, we 
show that the gauge invariant dimension-two condensates
$\vev{A^hA^h}$ and $\vev{\bar\varphi\varphi-\bar\omega\omega}$
are generated by the GZ dynamics, leading to the finite mass terms 
present in the RGZ effective theory.  Final remarks are gathered in Sec. V.

\section{The BRST-invariant Gribov-Zwanziger framework}\label{sec:BRST-inv-GZ}

 According to the gauge principle, a crucial step towards a well-defined theory  at the quantum level is 
 the establishment of the BRST invariance of the action. Let us briefly review the BRST-invariant formulation of the 
Gribov-Zwanziger action, which will allow us to extend previous 
results in the Landau gauge to a family of other gauges continually 
connected to it. Following \cite{Delbourgo:1986wz,Lavelle:1995ty,
Capri:2016aqq}, 
we start by  introducing  a composite $SU(N)$ gauge field
\begin{eqnarray}\label{eq:def-Ah}
 (A^h)_\mu = h^\dagger A_\mu h +\frac{i}{g}h^\dagger\partial_\mu h,
\end{eqnarray}
where 
\begin{eqnarray}
 h(x)=\exp\left(ig\xi^a(x)T^a\right)
\end{eqnarray}
is a matrix valued field, with $T^a$ ($a=1,\dots,N^2-1$) being 
the generators of $SU(N)$ and the field $\xi$ is related to the 
well-known Stueckelberg field \cite{Lavelle:1995ty}. With such 
a definition, it is possible to define a consistent nilpotent 
set of BRST transformations such that $s(A^h)_\mu^a=0$, as 
described in more details in the Appendix.

An important property of the field $A^h$ is that, in the Landau 
gauge, it may be effectively replaced by the gauge field $A$ in
any perturbative calculation, since the Stueckelberg field propagator 
$\langle \xi^a(p)\xi^b(-p)\rangle$ vanishes in this 
case \cite{Capri:2016aqq}.  In addition, the BRST invariant field $A^h$ field can be employed to 
obtain a BRST invariant expression for Zwanziger's horizon 
function  \cite{Capri:2015ixa,Capri:2016aqq}, namely
\begin{eqnarray}\label{eq:horizon-action-Ah}
 S'_h = \int\,d^dx\left[\bar\varphi_\mu^{ab}[{\cal M}(A^h)]^{ac} \varphi_\mu^{cb} - \bar\omega_\mu^{ab}[{\cal M}(A^h)]^{ac}\omega_\mu^{cb} 
 + ig\gamma^2f^{abc}(A^h)_\mu^a(\varphi^{bc}_\mu + \bar\varphi_\mu^{bc})\right],
\end{eqnarray}
 whose sole difference with respect to  (\ref{eq:horizon-funct-landau}) is 
the replacement $A\rightarrow A^h$.  Due to the properties of the field $A^h$, both expressions 
\eqref{eq:horizon-action} and \eqref{eq:horizon-action-Ah} turn out to be equivalent in the Landau gauge. 
Given the BRST invariance 
of $A^h$, it is consistent to define the auxiliary fields 
$\bar\varphi$, $\varphi$, $\bar\omega$, and $\omega$ as BRST 
singlets, so that the new horizon function (\ref{eq:horizon-action-Ah})
is itself BRST invariant. 

This now allows us to  generalize the GZ theory in other gauges different from the Landau 
gauge, {\it i.e.}: the linear covariant gauge 
\cite{Capri:2015ixa,Capri:2015nzw,Fiorentini:2016rwx,Capri:2016gut}, 
and  the more general family of covariant, color-preserving 
gauge-fixing conditions introduced in \cite{Capri:2018ijg}. 

 Since the 
horizon function displays BRST invariance, it follows immediately that the new 
Gribov-Zwanziger action
\begin{eqnarray}
 S'_{GZ} = S_{FP} + S'_h
\end{eqnarray}
is BRST invariant. As a consequence, correlation functions of 
BRST-invariant operators are independent of gauge parameters,
as discussed in \cite{Capri:2018ijg}. In particular, we can 
consider the class of Lorentz and color invariant gauges 
known as the $R_\xi$ gauges \cite{tHooft:1981bkw}. Its gauge 
fixing action is given by
\begin{eqnarray}\label{eq:gauge-fixing}
R_{\xi} & = & s\int d^{4}x\overline{c}^{a}\left(\partial_{\mu}A_{\mu}^{a}-\mu^{2}\xi^{a}-i\frac{\alpha}{2}b^{a}\right)\nonumber\\
 & = & \int d^{4}x\left(ib^{a}\partial_{\mu}A_{\mu}^{a}-\overline{c}^{a}{\cal M}^{ab}(A)c^{b}-i\mu^{2}b^{a}\xi^{a}
 +\mu^{2}\overline{c}^{a}g^{ab}\left(\xi\right)c^{b}+\frac{\alpha}{2}b^{a}b^{a}\right) \;, 
\end{eqnarray}
where $(\mu^2, \alpha)$ are gauge parameters and ${\cal M}^{ab}(A)$ is  the Faddeev-Popov operator in the $R_\xi$ gauge
\begin{eqnarray}
 {\cal M}^{ab}(A) = -\partial_\mu\,D^{ab}_\mu(A),
\end{eqnarray}
where 
\begin{eqnarray}
D^{ab}_\mu(A) = \delta^{ab}\partial_\mu - gf^{abc}A_\mu^c
\end{eqnarray}
is the covariant derivative in the adjoint representation of $SU(N)$.
The field functional $g^{ab}(\xi)$ is defined in eq. (\ref{eq:s-xi}) 
in the Appendix \ref{app:brst} and is derived from the BRST transformation 
of the Stueckelberg field $\xi$. 

Note that the $\mu\rightarrow0$ limit corresponds to the linear covariant 
gauge, in which $\alpha\rightarrow0$ is the Landau gauge. The massive 
parameter $\mu$ (roughly a ghost mass) works as an infrared  regulator  
for the Stueckelberg field $\xi$. This feature is especially useful 
for $\alpha\not=0$. For the Landau gauge, 
the Stueckelberg field $\xi$ decouples, since all its propagators vanish
\cite{Capri:2016aqq} and no infrared problems involving this auxiliary 
field appear.

Putting all ingredients in a single action, we obtain the 
BRST-invariant Gribov-Zwanziger action in the $R_\xi$ gauge
\begin{eqnarray}\label{eq:GZ-action}
S_{GZ} & = & \int d^{d}x\frac{1}{4}F_{\mu\nu}F_{\mu\nu}+R_{\xi}
+\int d^{d}x\left(\overline{\varphi}_{\mu}^{ac}
{\cal M}^{ab}(A^h)\varphi_{\mu}^{bc}-\overline{\omega}_{\mu}^{ac}
{\cal M}^{ab}(A^h)\omega_{\mu}^{bc}\right)+
 g\gamma^{2}f^{abc}\int d^{4}x{(A^h)}_{\mu}^{a}\left(\varphi_{\mu}^{bc}-\overline{\varphi}_{\mu}^{bc}\right)+\nonumber\\
 &  & +\int d^{4}x\left(i\tau^{a}\partial_{\mu}(A^h)_{\mu}^{a}+\overline{\eta}^{a}
 {\cal M}^{ab}(A^h)\eta^{b}\right).
\end{eqnarray}

It is important to point out here that the  expression above, eq.\eqref{eq:GZ-action}, is BRST invariant. As 
such, the Gribov parameter $\gamma^2$ acquires a 
potential physical meaning, being associated to a nontrivial BRST-invariant quantity, as expressed  by 
\begin{equation} 
\frac{\partial S_{GZ}}{\partial \gamma^2} = gf^{abc}\int d^{4}x{(A^h)}_{\mu}^{a}
\left(
\varphi_{\mu}^{bc}-\overline{\varphi}_{\mu}^{bc}
\right) 
\neq s{\hat \Delta} \;, \label{ng}
\end{equation} 
for any local field polynomial $\Delta$. Expression \eqref{ng} ensures that $\gamma^2$ is independent of the gauge parameters $(\alpha, \mu^2)$ entering the gauge $R_\xi$ condition. 

Finally, let us point out that the auxiliary field $\tau$ in expression \eqref{eq:GZ-action} 
is necessary to 
ensure the transversality of $A^h$, a feature that is crucial for the 
renormalizability of the GZ action (\ref{eq:GZ-action}) 
\cite{Capri:2017bfd}. Also, the new couple of ghosts $(\eta, \bar{\eta})$ takes 
into account the Jacobian arising from the integration over $\tau$ \cite{Capri:2017bfd}.

\section{The vacuum energy at leading order in the presence of constant sources}
\label{sec:vac-energy}

In order to study the dynamical generation of the condensates $\left\langle{A^hA^h}\right\rangle$ and 
$\left\langle{\bar\varphi\varphi - \bar\omega\omega}\right\rangle$, we introduce the corresponding operators 
in the GZ action, through constant sources $m^2$ and $M^2$, so that
\begin{eqnarray}\label{eq:GZ-action-with-sources}
S & = & S_{GZ}+\frac{m^{2}}{2}\int d^{4}x(A^h)^{a}_{\mu}(A^h)_{\mu}^{a}
+M^{2}\int d^{4}x\left(\overline{\varphi}_{\mu}^{ab}\varphi_{\mu}^{ab}-\overline{\omega}_{\mu}^{ab}\omega_{\mu}^{ab}\right).
\end{eqnarray}
In the presence of the sources $m^2$ and $M^2$, the partition function 
of the theory reads
\begin{eqnarray}
Z & = & \int\left[D\vec\phi\right]e^{-S} = e^{-V{\cal E}_{vac}},
\end{eqnarray}
where $[D\vec\phi]=[DA][Db][D\bar c][Dc][D\bar\varphi][D\varphi][D\bar\omega][D\omega]
[D\tau][D\xi][D\bar\eta][D\eta]$ 
represents the functional measure including all fields present in the action
and
${\cal E}_{vac}$ is the energy density of the vacuum state 
of the GZ theory, in the presence of the source terms, as in 
eq. (\ref{eq:GZ-action-with-sources}). Therefore,
\begin{eqnarray}\label{eq:vacuum-energy}
 {\cal E}_{vac} = -\frac{1}{V}\log Z = 
 \frac{1}{V}\log\int\left[D\vec\phi\right]e^{-S}.
\end{eqnarray}

In order to show that nonzero condensates appear already at 
leading order, let us consider the quadratic terms in the action 
(\ref{eq:GZ-action}), 
\begin{eqnarray}\label{eq:quadratic-action}
S_{quadr.} & = & \int d^{4}x
\left(\frac{1}{4}\left(\partial_{\mu}A_{\nu}^{a}-\partial_{\nu}A_{\mu}^{a}\right)^{2}
-\overline{\varphi}_{\mu}^{ab}\partial^{2}\varphi_{\mu}^{ab}
+\overline{\omega}_{\mu}^{ab}\partial^{2}\omega_{\mu}^{ab}\right.\nonumber\\
 &  & +g\gamma^{2}\left(A_{\mu}^{a}
 -\partial_{\mu}\xi^{a}\right)f^{abc}\left(\varphi_{\mu}^{bc}-\overline{\varphi}_{\mu}^{bc}\right)\nonumber\\
 &  & +i\tau^{a}\left(\partial_{\mu}A_{\mu}^{a}-
 \partial^{2}\xi^{a}\right)+\overline{\eta}^{a}\partial^{2}\eta^{a}\nonumber\\
 &  & +ib^{a}\partial_{\mu}A_{\mu}^{a}+\frac{\alpha}{2}b^{a}b^{a}-
 i\mu^{2}b^{a}\xi^{a}+\overline{c}^{a}\partial^{2}c^{a}-\mu^{2}\overline{c}^{a}c^{a}\nonumber\\
 &  & \left.+\frac{m^{2}}{2}\left(A_{\mu}^{a}-\partial_{\mu}\xi^{a}\right)\left(A_{\mu}^{a}-
 \partial_{\mu}\xi^{a}\right)+M^{2}\left(\overline{\varphi}_{\mu}^{ab}\varphi_{\mu}^{ab}-
 \overline{\omega}_{\mu}^{ab}\omega_{\mu}^{ab}\right)\right).
\end{eqnarray}

Within the quadratic approximation (\ref{eq:quadratic-action}), 
the vacuum energy (\ref{eq:vacuum-energy}) can be calculated 
by simply integrating out the fields iteratively using the standard 
formula \cite{Peskin:1995ev}
\begin{eqnarray}
 \int[D\phi]\exp\left[-\int\,d^dx\,\frac12\left(\phi\cdot \hat A\cdot \phi\right) 
 + \int d^dx\,J\cdot\phi\right] = (\det \hat A)^{-1/2}
 \exp\left(\frac12\int\,d^dx\,d^dy\,J(x)G_A(x,y)J(y)\right),
\end{eqnarray}
where $G_A$ is the Green's function of the operator $\hat A$.

It is convenient to start with the auxiliary fields $\tau$, $\bar\eta$ 
and $\eta$, which are related to the constraint on the transversality 
of $A^h$. Next, one may integrate out the Stueckelberg field $\xi$
and then Zwanziger's auxiliary fields $\bar\varphi$, $\varphi$, $\bar\omega$ 
and $\omega$, followed by the gauge fixing fields $b$, $\bar c$ and $c$.
At the end of this procedure, we are left with an integral over the 
gauge field $A$
\begin{eqnarray}\label{eq:partition-function-A}
 Z = \int[DA]\det(\alpha)^{-1/2}\det(-\partial^2+\mu^2)e^{-S_A},
\end{eqnarray}
where
\begin{eqnarray}\label{eq:gluon-action}
 S_A = \frac12\int\,d^dx\,A_\mu^a\delta^{ab}
 \left[\left(-\partial^2+m^2+\frac{2g^2N_c\gamma^4}{-\partial^2+M^2}\right)P_{\mu\nu} -\frac{1}{\alpha}\left(1+\frac{\mu^2}{-\partial^2}\right)^2\partial_\mu\partial_\nu \right]A_\nu^b
\end{eqnarray}
with the transverse projector
\begin{eqnarray}
 P_{\mu\nu} = \delta_{\mu\nu}-\frac{\partial_\mu\partial_\nu}{\partial^2}.
\end{eqnarray}

After the final integration in the gluon field, the gauge-dependent
terms present in the longitudinal part of the gluon action
(\ref{eq:gluon-action}) precisely cancel the determinants in 
(\ref{eq:partition-function-A}) and the vacuum energy density is, 
in the quadratic approximation,
\begin{eqnarray}\label{eq:vacuum-energy_quad}
 {\cal E}_{vac}^{(quad)} = \frac{N_c^2-1}{2}(d-1)\int\frac{d^dp}{(2\pi)^d}
 \log\left(p^2+m^2+\frac{\Lambda^4}{p^2+M^2}\right) - 
 \frac{N_c^2-1}{2}\int\frac{d^dp}{(2\pi)^d}\log p^2,
\end{eqnarray}
where $\Lambda^4 = 2g^2N_c\gamma^4$. Let us remark that the independence of
(\ref{eq:vacuum-energy_quad}) with respect to the gauge parameters $\alpha$ and 
$\mu^2$ is an explicit consequence of the exact BRST invariance of the present 
formulation of the Gribov-Zwanziger framework (\ref{eq:GZ-action}).
 Note that this property would not be true for the original
Gribov-Zwanziger framework, in which the BRST symmetry is softly broken.
This is the first important feature of the proposed action
(\ref{eq:GZ-action}) that we wish to show in this paper.

\section{Two gauge invariant condensates and the instability of the GZ action}
\label{sec:condensates}

Now that we have calculated the vacuum energy density in the quadratic 
approximation, we may proceed to calculate the condensates
\begin{eqnarray}
 \vev{(A^h)^a_\mu(A^h)^a_\mu} &=& \frac{1}{Z}\int[D\vec{\Phi}](A^h)^a_\mu(A^h)^a_\mu 
 e^{-S_{GZ}}
\end{eqnarray}
and
\begin{eqnarray}
 \vev{\bar\varphi^{ab}_\mu\varphi^{ab}_\mu - \bar\omega^{ab}_\mu\omega^{ab}_\mu} &=& \frac{1}{Z}\int[D\vec{\Phi}](\bar\varphi^{ab}_\mu\varphi^{ab}_\mu - \bar\omega^{ab}_\mu\omega^{ab}_\mu) e^{-S_{GZ}}.
\end{eqnarray}

Note that these condensates are calculated within  the BRST invariant formulation of the Gribov-Zwanziger 
theory, hence the action $S_{GZ}$, eq. (\ref{eq:GZ-action}), in the 
definitions above. An important property of both condensates is that 
they correspond to expectation values of BRST invariant operators, as
briefly reviewed in Appendix \ref{app:brst}. They can furthermore be 
expressed in terms of the vacuum energy as
\begin{eqnarray}\label{eq:A2-cond}
\left\langle (A^h)_{\mu}^{a}(A^h)_{\mu}^{a}\right\rangle  & = & 2\left.\frac{\partial{\cal E}_{vac}}{\partial m^{2}}\right|_{m^{2}=M^{2}=0}
\end{eqnarray}
and
\begin{eqnarray}\label{eq:phi-omega-cond}
\left\langle \overline{\varphi}^{ab}_\mu\varphi^{ab}_\mu-
\overline{\omega}^{ab}_\mu\omega^{ab}_\mu\right\rangle  & = & \left.\frac{\partial{\cal E}_{vac}}{\partial M^{2}}\right|_{m^{2}=M^{2}=0}.
\end{eqnarray}

Note that the sources $m^2$ and $M^2$ are taken to zero at the end of 
the calculation. This is analogous to considering the appearance of 
a spontaneous magnetization in a spin model: first one considers the 
interaction of the magnetic moments in the presence of some external 
magnetic field and then, at the last step of the calculation, the 
external field is taken to zero. A resulting nonzero value of the 
condensates $\vev{A^hA^h}$ or $\vev{\bar\varphi\varphi - \bar\omega\omega}$
starting from the GZ action (\ref{eq:GZ-action}) is somewhat analogous 
to a spontaneous magnetization in a spin model \footnote{With an important 
difference that the operators which condense in our case are not 
elementary fields that enter the lagrangian, as would be the case for
spontaneous magnetization in a spin model.}.

Even though our calculation already shows that the condensates 
$\vev{A^hA^h}$ and $\vev{\bar\varphi\varphi - \bar\omega\omega}$
are nonzero, a fuller calculation would  require the construction of  
effective potential for the expectation values of these two operators, a task which can be 
faced by means of the Local Composite Operator (LCO) 
formalism  \cite{Verschelde:2001ia,Knecht:2001cc},  see \cite{Dudal:2003vv,Browne:2003uv,Dudal:2005na,Dudal:2011gd} for 
previous attempts in the context of the RGZ theory in the Landau gauge  and \cite{Dudal:2018ctc}
 for recent developments.

From the expression (\ref{eq:vacuum-energy_quad}) for the vacuum energy and 
formulas (\ref{eq:A2-cond}) and (\ref{eq:phi-omega-cond}), the 
condensates can be immediately calculated, leading to
\begin{eqnarray}\label{eq:A2-cond-2}
 \vev{(A^h)_{\mu}^{a}(A^h)_{\mu}^{a}} &=&\frac{(N_c^2-1)(d-1)}2\int\frac{d^dp}{(2\pi)^d}\frac{1}{p^2+\frac{\Lambda^4}{p^2}}\nonumber\\
 &=& -\frac{(N_c^2-1)(d-1)}2 I_d(\Lambda)
\end{eqnarray}
and
\begin{eqnarray}\label{eq:phi-omega-cond-2}
 \vev{\overline{\varphi}^{ab}_\mu\varphi^{ab}_\mu-
\overline{\omega}^{ab}_\mu\omega^{ab}_\mu} 
&=&\frac{(N_c^2-1)(d-1)}2\left.\int\frac{d^dp}{(2\pi)^d}\frac{p^2}{p^4+\Lambda^4}\frac{-\Lambda^4}{(p^2+M^2)^2}\right|_{M=0}\nonumber\\
 &=&-\frac{(N_c^2-1)(d-1)}2 I_d(\Lambda)
\end{eqnarray}
where we used the decomposition 
\begin{eqnarray}
 \frac{1}{p^2+\frac{\Lambda^4}{p^2}} = \frac{1}{p^2} - \frac{\Lambda^4}{p^2(p^4+\Lambda^4)}
\end{eqnarray}
and the dimensional regularization formula
\begin{eqnarray}
 \int\frac{d^dp}{(2\pi)^d}\frac1{p^2} = 0,
\end{eqnarray}
which is valid for $d>2$. For the final expressions, we also used
the results
\begin{align}\label{eq:integral}
I_d(\Lambda):=\int\frac{d^dp}{(2\pi)^d}\frac{\Lambda^4}{p^2(p^4+\Lambda^4)} 
& = \frac{\Lambda^2}{32\pi} &(d=4)\nonumber\\
& = \frac{\sqrt{2}\Lambda}{8\pi} &(d=3)\nonumber\\
& \longrightarrow\infty &(d=2).
\end{align}
Note that, for $d=2$, the integral 
(\ref{eq:integral}) is IR divergent, so that the condensates do not show up, as could be anticipated by the Mermin-Wagner-Coleman theorem 
\cite{MerminWagner,Coleman:1973ci}. In $d=4$ and in $d=3$,  these leading-order results for the condensates 
vanish if 
one does not consider the restriction to the Gribov region, i.e., 
if $\gamma=0$ (or, equivalently, $\Lambda=0$). It is important to note that it has been shown in pure Yang-Mills theories in Landau gauge that the $\langle A^2 \rangle$ condensate is generated by the non-Abelian interaction, even in the absence of the Gribov horizon \cite{Dudal:2003vv,Browne:2003uv}. This finding is compatible with our quadratic analysis because in our aproximation we neglect the direct effect of interactions, considering only the presence of the nonperturbative background brought about by the constraint of the Gribov horizon condition. A full fledged effective potential calculation would therefore reveal that the $\vev{(A^h)_{\mu}^{a}(A^h)_{\mu}^{a}}$ condensate has a nonzero limit as the Gribov parameter vanishes. 

Let us finally note that the Gribov parameter is independent 
of the gauge parameters, being thus allowed to enter explicitly findings for physical observables.  This can be immediately seen from 
the defining equation of $\gamma$, the gap equation
\cite{Gribov:1977wm}
\begin{eqnarray}
 \frac{\partial{\cal E}_{vac}}{\partial\gamma^2} = 0.
\end{eqnarray}
 where ${\cal E}_{vac}$ is the vacuum energy (\ref{eq:vacuum-energy_quad}),  which is 
explicitly independent of the gauge parameters  $\alpha$ and $\mu$. 

 Due to the BRST invariance, it turns out that 
the condensates $\vev{(A^h)_{\mu}^{a}(A^h)_{\mu}^{a}}$ and 
$\vev{\overline{\varphi}^{ab}_\mu\varphi^{ab}_\mu-
\overline{\omega}^{ab}_\mu\omega^{ab}_\mu} $, (\ref{eq:A2-cond-2}) and (\ref{eq:phi-omega-cond-2}), are 
also independent of the gauge parameters $\alpha$ and $\mu$ appearing in the gauge condition. 
This is a direct 
consequence of the  BRST-invariant formulation 
of the Gribov-Zwanziger theory put forward in 
\cite{Capri:2015ixa,Capri:2015nzw,Capri:2016aqq}. 
It is also interesting to note that, due to the 
independence of the energy 
density on the gauge parameters, if the condensation of 
these operators happens for some value of $\alpha$ and $\mu$
(for example, for the Landau gauge), it will happen for 
any values of the gauge parameters. This property indicate that
indeed these condensates may have a genuine physical meaning.
Furthermore, the RGZ parameters
$m^2$ and $M^2$ related to the condensates do not depend on 
the gauge parameters and are thus given by their Landau gauge 
values. Finally, although the gluon propagator 
$\vev{A_\mu^a(p)A_\nu^b(-p)}$ is gauge dependent, the two-point 
correlation point of the $A^h$ composite field is gauge independent,
being equal to the Landau gauge gluon propagator. Thus, it is 
possible to use this idea to predict how the gluon propagator deviates from 
the Landau gauge value as one considers nonzero values for 
the gauge parameters $\alpha$ and $\mu$ for some generic $R_\xi$ 
gauges. 

\section{Final remarks}

In this work, we have extended previous discussions on the generation of 
dimension two condensates in the Landau gauge Gribov-Zwanziger framework 
to a two-parameter family of gauges, namely, the $R_\xi$ gauges, parametrized
by the parameters $\alpha$ and $\mu$. In the limit $\mu\rightarrow0$, 
one recovers the Linear Covariant Gauge, so that with $(\mu,\alpha)\rightarrow0$ 
one achieves the Landau gauge. 
We have explicitly shown that the instability of the GZ action observed in the Landau gauge is actually a universal property in this large class of gauges, suggesting a genuine physical meaning for the refinement of the GZ action in the infrared regime.
In particular, we have 
shown that the GZ action is unstable with respect to the formation of 
the BRST invariant condensates $\vev{(A^h)_{\mu}^{a}(A^h)_{\mu}^{a}}$ and 
$\vev{\overline{\varphi}^{ab}_\mu\varphi^{ab}_\mu-
\overline{\omega}^{ab}_\mu\omega^{ab}_\mu}$
, which were computed at leading order. These 
condensates are proportional to the Gribov parameter $\gamma^2$, which 
is also independent of the choice of  the gauge parameters contained in the 
 $R_\xi$ gauges. 
 
 Such gauge independence  reinforces  the fact that both 
Gribov parameter and  condensates enter the correlation functions of physical operators, {\it i.e.} 
the correlation functions of local gauge invariant quantity as, for example, the glueball 
spectrum (cf. e.g. \cite{Dudal:2013wja,Dudal:2010cd}). The results presented here are therefore a step forward in proving that the RGZ framework can provide a physically meaningful description of the infrared regime of Yang-Mills theories.

 An interesting point raised by these developments is whether one can probe directly these gauge-invariant dimension-two condensates on the lattice. For that one would need to write these nonlocal operators in terms of lattice variables, a task which is not straightforward. Nevertheless, as we have emphasized, the current description of the RGZ formalism in $R_{\xi}$ gauges 
 implies 
that the numerical results 
 for expectation values of BRST invariant operators  must be identical to those in the Landau gauge, which is already accessible by lattice methods. Indeed there are indications of a nonzero $\vev{A^2}$ in lattice Yang-Mills simulations for the Landau gauge \cite{Pene:2011kg,Chernodub:2008kf} and we expect to report on direct comparisons between the RGZ predictions and the lattice results in the future.


\section*{Acknowledgements}

The authors would like to thank the Brazilian agencies 
CNPq and FAPERJ for financial support. This paper is  also  
part  of  the project  INCT-FNA  Process  No. 464898/2014-5. 
B.~W.~M. is supported by CNPq project Universal (grant 431796/2016-5) 
and FAPERJ (grant E-26/202.649/2018).
This study was financed in part by the Coordena\c c\~ao de 
Aperfei\c coamento de Pessoal de N\'\i vel Superior $-$ Brasil (CAPES) $-$ Financial Code 001 (M. N. F.).

\appendix

\section{The BRST transformations and the gauge invariance of the GZ action}
\label{app:brst}

For completeness, let us introduce the BRST transformations of the 
fields present in the GZ action (\ref{eq:GZ-action}) in order to 
show that such an action (as well as its version with sources, 
eq. (\ref{eq:GZ-action-with-sources})) is BRST invariant. The BRST
transformations of the fields are 
\begin{eqnarray}\label{eq:brst-YM}
 sA_\mu^a &=& -D_\mu^{ab}c^b\nonumber\\
 sc^a &=& \frac{g}{2}f^{abc}c^bc^c\nonumber\\
 s\bar c^a &=& ib^a\nonumber\\
 sb^a &=& 0,
\end{eqnarray}
as in standard perturbative Yang-Mills theories, supplemented by 
the trivial transformations of the auxiliary fields
\begin{eqnarray}\label{eq:s-auxiliary}
 s\Phi = 0,
\end{eqnarray}
where $\Phi$ is any of the fields $\bar\varphi,\varphi,\bar\omega,\omega,\tau,
\bar\eta,\eta$. The field $A^h$ is, by construction, BRST invariant, so that
\begin{eqnarray}\label{eq:sAh}
 s(A^h)^a_\mu = 0.
\end{eqnarray}
From the definition (\ref{eq:def-Ah}) of $A^h$, the condition (\ref{eq:sAh}) 
leads to the particular BRST transformation of the Stueckelberg 
field $\xi$ \cite{Capri:2016aqq}
\begin{eqnarray}
 s\xi^a = g^{ab}c^b,
\end{eqnarray}
where
\begin{eqnarray}\label{eq:s-xi}
 g^{ab}(\xi) = -\delta^{ab} + \frac{g}{2}f^{abc}\xi^c -
 \frac{g^2}{12}f^{amr}f^{mbq}\xi^q\xi^r + {\cal O}(\xi^3).
\end{eqnarray}

As is well-known, the Faddeev-Popov action 
\begin{eqnarray}
 S_{FP} = \int\,d^dx\,\frac{1}{4}F_{\mu\nu}F_{\mu\nu} + R_\xi,
\end{eqnarray}
is invariant under the BRST transformations (\ref{eq:brst-YM}), 
where $R_\xi$ is the gauge-fixing action in (\ref{eq:gauge-fixing}).
Furthermore, with the BRST transformations (\ref{eq:s-auxiliary}) and
(\ref{eq:sAh}) one can immediately check that the Gribov-Zwanziger action
(\ref{eq:GZ-action}) is invariant under BRST transformations, as 
advertised. 

Finally, note that the addition of the BRST invariant sources
$\vev{(A^h)_{\mu}^{a}(A^h)_{\mu}^{a}}$ and 
$\vev{\overline{\varphi}^{ab}_\mu\varphi^{ab}_\mu-
\overline{\omega}^{ab}_\mu\omega^{ab}_\mu} $  
does not spoil the BRST invariance of the action 
(\ref{eq:GZ-action-with-sources}).

\bibliography{RGZ-bibliography}

\begin{thebibliography}{53}%
\makeatletter
\providecommand \@ifxundefined [1]{%
 \@ifx{#1\undefined}
}%
\providecommand \@ifnum [1]{%
 \ifnum #1\expandafter \@firstoftwo
 \else \expandafter \@secondoftwo
 \fi
}%
\providecommand \@ifx [1]{%
 \ifx #1\expandafter \@firstoftwo
 \else \expandafter \@secondoftwo
 \fi
}%
\providecommand \natexlab [1]{#1}%
\providecommand \enquote  [1]{``#1''}%
\providecommand \bibnamefont  [1]{#1}%
\providecommand \bibfnamefont [1]{#1}%
\providecommand \citenamefont [1]{#1}%
\providecommand \href@noop [0]{\@secondoftwo}%
\providecommand \href [0]{\begingroup \@sanitize@url \@href}%
\providecommand \@href[1]{\@@startlink{#1}\@@href}%
\providecommand \@@href[1]{\endgroup#1\@@endlink}%
\providecommand \@sanitize@url [0]{\catcode `\\12\catcode `\$12\catcode
  `\&12\catcode `\#12\catcode `\^12\catcode `\_12\catcode `\%12\relax}%
\providecommand \@@startlink[1]{}%
\providecommand \@@endlink[0]{}%
\providecommand \url  [0]{\begingroup\@sanitize@url \@url }%
\providecommand \@url [1]{\endgroup\@href {#1}{\urlprefix }}%
\providecommand \urlprefix  [0]{URL }%
\providecommand \Eprint [0]{\href }%
\providecommand \doibase [0]{http://dx.doi.org/}%
\providecommand \selectlanguage [0]{\@gobble}%
\providecommand \bibinfo  [0]{\@secondoftwo}%
\providecommand \bibfield  [0]{\@secondoftwo}%
\providecommand \translation [1]{[#1]}%
\providecommand \BibitemOpen [0]{}%
\providecommand \bibitemStop [0]{}%
\providecommand \bibitemNoStop [0]{.\EOS\space}%
\providecommand \EOS [0]{\spacefactor3000\relax}%
\providecommand \BibitemShut  [1]{\csname bibitem#1\endcsname}%
\let\auto@bib@innerbib\@empty
\bibitem [{\citenamefont {Faddeev}\ and\ \citenamefont
  {Popov}(1967)}]{Faddeev:1967fc}%
  \BibitemOpen
  \bibfield  {author} {\bibinfo {author} {\bibfnamefont {L.~D.}\ \bibnamefont
  {Faddeev}}\ and\ \bibinfo {author} {\bibfnamefont {V.~N.}\ \bibnamefont
  {Popov}},\ }\href {\doibase 10.1016/0370-2693(67)90067-6} {\bibfield
  {journal} {\bibinfo  {journal} {Phys. Lett.}\ }\textbf {\bibinfo {volume}
  {B25}},\ \bibinfo {pages} {29} (\bibinfo {year} {1967})},\ \bibinfo {note}
  {[,325(1967)]}\BibitemShut {NoStop}%
\bibitem [{\citenamefont {Peskin}\ and\ \citenamefont
  {Schroeder}(1995)}]{Peskin:1995ev}%
  \BibitemOpen
  \bibfield  {author} {\bibinfo {author} {\bibfnamefont {M.~E.}\ \bibnamefont
  {Peskin}}\ and\ \bibinfo {author} {\bibfnamefont {D.~V.}\ \bibnamefont
  {Schroeder}},\ }\href {http://www.slac.stanford.edu/~mpeskin/QFT.html} {\emph
  {\bibinfo {title} {{An Introduction to quantum field theory}}}}\ (\bibinfo
  {publisher} {Addison-Wesley},\ \bibinfo {address} {Reading, USA},\ \bibinfo
  {year} {1995})\BibitemShut {NoStop}%
\bibitem [{\citenamefont {Weinberg}(2005)}]{Weinberg:1995mt}%
  \BibitemOpen
  \bibfield  {author} {\bibinfo {author} {\bibfnamefont {S.}~\bibnamefont
  {Weinberg}},\ }\href@noop {} {\emph {\bibinfo {title} {{The Quantum theory of
  fields. Vol. 1: Foundations}}}}\ (\bibinfo  {publisher} {Cambridge University
  Press},\ \bibinfo {year} {2005})\BibitemShut {NoStop}%
\bibitem [{\citenamefont {Zwanziger}(2005)}]{Zwanziger:2004np}%
  \BibitemOpen
  \bibfield  {author} {\bibinfo {author} {\bibfnamefont {D.}~\bibnamefont
  {Zwanziger}},\ }\href {\doibase 10.1103/PhysRevLett.94.182301} {\bibfield
  {journal} {\bibinfo  {journal} {Phys. Rev. Lett.}\ }\textbf {\bibinfo
  {volume} {94}},\ \bibinfo {pages} {182301} (\bibinfo {year} {2005})},\
  \Eprint {http://arxiv.org/abs/hep-ph/0407103} {arXiv:hep-ph/0407103 [hep-ph]}
  \BibitemShut {NoStop}%
\bibitem [{\citenamefont {Guimaraes}\ \emph {et~al.}(2015)\citenamefont
  {Guimaraes}, \citenamefont {Mintz},\ and\ \citenamefont
  {Sorella}}]{Guimaraes:2015bra}%
  \BibitemOpen
  \bibfield  {author} {\bibinfo {author} {\bibfnamefont {M.~S.}\ \bibnamefont
  {Guimaraes}}, \bibinfo {author} {\bibfnamefont {B.~W.}\ \bibnamefont
  {Mintz}}, \ and\ \bibinfo {author} {\bibfnamefont {S.~P.}\ \bibnamefont
  {Sorella}},\ }\href {\doibase 10.1103/PhysRevD.91.121701} {\bibfield
  {journal} {\bibinfo  {journal} {Phys. Rev.}\ }\textbf {\bibinfo {volume}
  {D91}},\ \bibinfo {pages} {121701} (\bibinfo {year} {2015})},\ \Eprint
  {http://arxiv.org/abs/1503.03120} {arXiv:1503.03120 [hep-th]} \BibitemShut
  {NoStop}%
\bibitem [{\citenamefont {Campagnari}\ \emph {et~al.}(2016)\citenamefont
  {Campagnari}, \citenamefont {Ebadati}, \citenamefont {Reinhardt},\ and\
  \citenamefont {Vastag}}]{Campagnari:2016wlt}%
  \BibitemOpen
  \bibfield  {author} {\bibinfo {author} {\bibfnamefont {D.~R.}\ \bibnamefont
  {Campagnari}}, \bibinfo {author} {\bibfnamefont {E.}~\bibnamefont {Ebadati}},
  \bibinfo {author} {\bibfnamefont {H.}~\bibnamefont {Reinhardt}}, \ and\
  \bibinfo {author} {\bibfnamefont {P.}~\bibnamefont {Vastag}},\ }\href
  {\doibase 10.1103/PhysRevD.94.074027} {\bibfield  {journal} {\bibinfo
  {journal} {Phys. Rev.}\ }\textbf {\bibinfo {volume} {D94}},\ \bibinfo {pages}
  {074027} (\bibinfo {year} {2016})},\ \Eprint
  {http://arxiv.org/abs/1608.06820} {arXiv:1608.06820 [hep-ph]} \BibitemShut
  {NoStop}%
\bibitem [{\citenamefont {'t~Hooft}(1981)}]{tHooft:1981bkw}%
  \BibitemOpen
  \bibfield  {author} {\bibinfo {author} {\bibfnamefont {G.}~\bibnamefont
  {'t~Hooft}},\ }\href {\doibase 10.1016/0550-3213(81)90442-9} {\bibfield
  {journal} {\bibinfo  {journal} {Nucl. Phys.}\ }\textbf {\bibinfo {volume}
  {B190}},\ \bibinfo {pages} {455} (\bibinfo {year} {1981})}\BibitemShut
  {NoStop}%
\bibitem [{\citenamefont {Kronfeld}\ \emph
  {et~al.}(1987{\natexlab{a}})\citenamefont {Kronfeld}, \citenamefont
  {Laursen}, \citenamefont {Schierholz},\ and\ \citenamefont
  {Wiese}}]{kronfeld:1987ri}%
  \BibitemOpen
  \bibfield  {author} {\bibinfo {author} {\bibfnamefont {A.~S.}\ \bibnamefont
  {Kronfeld}}, \bibinfo {author} {\bibfnamefont {M.~L.}\ \bibnamefont
  {Laursen}}, \bibinfo {author} {\bibfnamefont {G.}~\bibnamefont {Schierholz}},
  \ and\ \bibinfo {author} {\bibfnamefont {U.~J.}\ \bibnamefont {Wiese}},\
  }\href {\doibase 10.1016/0370-2693(87)90910-5} {\bibfield  {journal}
  {\bibinfo  {journal} {Phys. Lett.}\ }\textbf {\bibinfo {volume} {B198}},\
  \bibinfo {pages} {516} (\bibinfo {year} {1987}{\natexlab{a}})}\BibitemShut
  {NoStop}%
\bibitem [{\citenamefont {Kronfeld}\ \emph
  {et~al.}(1987{\natexlab{b}})\citenamefont {Kronfeld}, \citenamefont
  {Schierholz},\ and\ \citenamefont {Wiese}}]{Kronfeld:1987vd}%
  \BibitemOpen
  \bibfield  {author} {\bibinfo {author} {\bibfnamefont {A.~S.}\ \bibnamefont
  {Kronfeld}}, \bibinfo {author} {\bibfnamefont {G.}~\bibnamefont
  {Schierholz}}, \ and\ \bibinfo {author} {\bibfnamefont {U.~J.}\ \bibnamefont
  {Wiese}},\ }\href {\doibase 10.1016/0550-3213(87)90080-0} {\bibfield
  {journal} {\bibinfo  {journal} {Nucl. Phys.}\ }\textbf {\bibinfo {volume}
  {B293}},\ \bibinfo {pages} {461} (\bibinfo {year}
  {1987}{\natexlab{b}})}\BibitemShut {NoStop}%
\bibitem [{\citenamefont {Gaigg}\ \emph {et~al.}(1990)\citenamefont {Gaigg},
  \citenamefont {Kummer},\ and\ \citenamefont {Schweda}}]{Gaigg:1990si}%
  \BibitemOpen
  \bibinfo {editor} {\bibfnamefont {P.}~\bibnamefont {Gaigg}}, \bibinfo
  {editor} {\bibfnamefont {W.}~\bibnamefont {Kummer}}, \ and\ \bibinfo {editor}
  {\bibfnamefont {M.}~\bibnamefont {Schweda}},\ eds.,\ \href {\doibase
  10.1007/BFb0015131} {\emph {\bibinfo {title} {{Physical and nonstandard
  gauges. Proceedings, Workshop, Vienna, Austria, September 19-23, 1989}}}},\
  Vol.\ \bibinfo {volume} {361}\ (\bibinfo {year} {1990})\BibitemShut {NoStop}%
\bibitem [{\citenamefont {Wilson}(1974)}]{Wilson:1974sk}%
  \BibitemOpen
  \bibfield  {author} {\bibinfo {author} {\bibfnamefont {K.~G.}\ \bibnamefont
  {Wilson}},\ }\href {\doibase 10.1103/PhysRevD.10.2445} {\bibfield  {journal}
  {\bibinfo  {journal} {Phys. Rev.}\ }\textbf {\bibinfo {volume} {D10}},\
  \bibinfo {pages} {2445} (\bibinfo {year} {1974})},\ \bibinfo {note}
  {[,319(1974)]}\BibitemShut {NoStop}%
\bibitem [{\citenamefont {Gribov}(1978)}]{Gribov:1977wm}%
  \BibitemOpen
  \bibfield  {author} {\bibinfo {author} {\bibfnamefont {V.~N.}\ \bibnamefont
  {Gribov}},\ }\href {\doibase 10.1016/0550-3213(78)90175-X} {\bibfield
  {journal} {\bibinfo  {journal} {Nucl. Phys.}\ }\textbf {\bibinfo {volume}
  {B139}},\ \bibinfo {pages} {1} (\bibinfo {year} {1978})}\BibitemShut
  {NoStop}%
\bibitem [{Note1()}]{Note1}%
  \BibitemOpen
  \bibinfo {note} {An analogous argument can also be cast in the other
  mentioned gauges.}\BibitemShut {Stop}%
\bibitem [{\citenamefont {Zwanziger}(1989)}]{Zwanziger:1989mf}%
  \BibitemOpen
  \bibfield  {author} {\bibinfo {author} {\bibfnamefont {D.}~\bibnamefont
  {Zwanziger}},\ }\href {\doibase 10.1016/0550-3213(89)90122-3} {\bibfield
  {journal} {\bibinfo  {journal} {Nucl. Phys.}\ }\textbf {\bibinfo {volume}
  {B323}},\ \bibinfo {pages} {513} (\bibinfo {year} {1989})}\BibitemShut
  {NoStop}%
\bibitem [{\citenamefont {Vandersickel}\ and\ \citenamefont
  {Zwanziger}(2012)}]{Vandersickel:2012tz}%
  \BibitemOpen
  \bibfield  {author} {\bibinfo {author} {\bibfnamefont {N.}~\bibnamefont
  {Vandersickel}}\ and\ \bibinfo {author} {\bibfnamefont {D.}~\bibnamefont
  {Zwanziger}},\ }\href {\doibase 10.1016/j.physrep.2012.07.003} {\bibfield
  {journal} {\bibinfo  {journal} {Phys. Rept.}\ }\textbf {\bibinfo {volume}
  {520}},\ \bibinfo {pages} {175} (\bibinfo {year} {2012})},\ \Eprint
  {http://arxiv.org/abs/1202.1491} {arXiv:1202.1491 [hep-th]} \BibitemShut
  {NoStop}%
\bibitem [{\citenamefont {Osterwalder}\ and\ \citenamefont
  {Schrader}(1973)}]{Osterwalder:1973dx}%
  \BibitemOpen
  \bibfield  {author} {\bibinfo {author} {\bibfnamefont {K.}~\bibnamefont
  {Osterwalder}}\ and\ \bibinfo {author} {\bibfnamefont {R.}~\bibnamefont
  {Schrader}},\ }\href {\doibase 10.1007/BF01645738} {\bibfield  {journal}
  {\bibinfo  {journal} {Commun. Math. Phys.}\ }\textbf {\bibinfo {volume}
  {31}},\ \bibinfo {pages} {83} (\bibinfo {year} {1973})}\BibitemShut {NoStop}%
\bibitem [{\citenamefont {Osterwalder}\ and\ \citenamefont
  {Schrader}(1975)}]{Osterwalder:1974tc}%
  \BibitemOpen
  \bibfield  {author} {\bibinfo {author} {\bibfnamefont {K.}~\bibnamefont
  {Osterwalder}}\ and\ \bibinfo {author} {\bibfnamefont {R.}~\bibnamefont
  {Schrader}},\ }\href {\doibase 10.1007/BF01608978} {\bibfield  {journal}
  {\bibinfo  {journal} {Commun. Math. Phys.}\ }\textbf {\bibinfo {volume}
  {42}},\ \bibinfo {pages} {281} (\bibinfo {year} {1975})}\BibitemShut
  {NoStop}%
\bibitem [{\citenamefont {Oehme}(1995)}]{Oehme:1994pv}%
  \BibitemOpen
  \bibfield  {author} {\bibinfo {author} {\bibfnamefont {R.}~\bibnamefont
  {Oehme}},\ }\bibfield  {booktitle} {\emph {\bibinfo {booktitle}
  {{Mathematical physics. Proceedings, 11th International Congress, Paris,
  France, July 18-22, 1994}}},\ }\href {\doibase 10.1142/S0217751X95000978}
  {\bibfield  {journal} {\bibinfo  {journal} {Int. J. Mod. Phys.}\ }\textbf
  {\bibinfo {volume} {A10}},\ \bibinfo {pages} {1995} (\bibinfo {year}
  {1995})},\ \Eprint {http://arxiv.org/abs/hep-th/9412040}
  {arXiv:hep-th/9412040 [hep-th]} \BibitemShut {NoStop}%
\bibitem [{\citenamefont {Nishijima}(1996)}]{Nishijima:1996ji}%
  \BibitemOpen
  \bibfield  {author} {\bibinfo {author} {\bibfnamefont {K.}~\bibnamefont
  {Nishijima}},\ }\href {\doibase 10.1007/BF01692238} {\bibfield  {journal}
  {\bibinfo  {journal} {Czech. J. Phys.}\ }\textbf {\bibinfo {volume} {46}},\
  \bibinfo {pages} {1} (\bibinfo {year} {1996})}\BibitemShut {NoStop}%
\bibitem [{\citenamefont {Alkofer}\ and\ \citenamefont {von
  Smekal}(2001)}]{Alkofer:2000wg}%
  \BibitemOpen
  \bibfield  {author} {\bibinfo {author} {\bibfnamefont {R.}~\bibnamefont
  {Alkofer}}\ and\ \bibinfo {author} {\bibfnamefont {L.}~\bibnamefont {von
  Smekal}},\ }\href {\doibase 10.1016/S0370-1573(01)00010-2} {\bibfield
  {journal} {\bibinfo  {journal} {Phys. Rept.}\ }\textbf {\bibinfo {volume}
  {353}},\ \bibinfo {pages} {281} (\bibinfo {year} {2001})},\ \Eprint
  {http://arxiv.org/abs/hep-ph/0007355} {arXiv:hep-ph/0007355 [hep-ph]}
  \BibitemShut {NoStop}%
\bibitem [{\citenamefont {Cucchieri}\ and\ \citenamefont
  {Mendes}(2008{\natexlab{a}})}]{Cucchieri:2007rg}%
  \BibitemOpen
  \bibfield  {author} {\bibinfo {author} {\bibfnamefont {A.}~\bibnamefont
  {Cucchieri}}\ and\ \bibinfo {author} {\bibfnamefont {T.}~\bibnamefont
  {Mendes}},\ }\href {\doibase 10.1103/PhysRevLett.100.241601} {\bibfield
  {journal} {\bibinfo  {journal} {Phys. Rev. Lett.}\ }\textbf {\bibinfo
  {volume} {100}},\ \bibinfo {pages} {241601} (\bibinfo {year}
  {2008}{\natexlab{a}})},\ \Eprint {http://arxiv.org/abs/0712.3517}
  {arXiv:0712.3517 [hep-lat]} \BibitemShut {NoStop}%
\bibitem [{\citenamefont {Fischer}\ \emph {et~al.}(2009)\citenamefont
  {Fischer}, \citenamefont {Maas},\ and\ \citenamefont
  {Pawlowski}}]{Fischer:2008uz}%
  \BibitemOpen
  \bibfield  {author} {\bibinfo {author} {\bibfnamefont {C.~S.}\ \bibnamefont
  {Fischer}}, \bibinfo {author} {\bibfnamefont {A.}~\bibnamefont {Maas}}, \
  and\ \bibinfo {author} {\bibfnamefont {J.~M.}\ \bibnamefont {Pawlowski}},\
  }\href {\doibase 10.1016/j.aop.2009.07.009} {\bibfield  {journal} {\bibinfo
  {journal} {Annals Phys.}\ }\textbf {\bibinfo {volume} {324}},\ \bibinfo
  {pages} {2408} (\bibinfo {year} {2009})},\ \Eprint
  {http://arxiv.org/abs/0810.1987} {arXiv:0810.1987 [hep-ph]} \BibitemShut
  {NoStop}%
\bibitem [{\citenamefont {Oliveira}\ and\ \citenamefont
  {Silva}(2009)}]{Oliveira:2008uf}%
  \BibitemOpen
  \bibfield  {author} {\bibinfo {author} {\bibfnamefont {O.}~\bibnamefont
  {Oliveira}}\ and\ \bibinfo {author} {\bibfnamefont {P.~J.}\ \bibnamefont
  {Silva}},\ }\href {\doibase 10.1103/PhysRevD.79.031501} {\bibfield  {journal}
  {\bibinfo  {journal} {Phys. Rev.}\ }\textbf {\bibinfo {volume} {D79}},\
  \bibinfo {pages} {031501} (\bibinfo {year} {2009})},\ \Eprint
  {http://arxiv.org/abs/0809.0258} {arXiv:0809.0258 [hep-lat]} \BibitemShut
  {NoStop}%
\bibitem [{\citenamefont {Cucchieri}\ and\ \citenamefont
  {Mendes}(2008{\natexlab{b}})}]{Cucchieri:2008fc}%
  \BibitemOpen
  \bibfield  {author} {\bibinfo {author} {\bibfnamefont {A.}~\bibnamefont
  {Cucchieri}}\ and\ \bibinfo {author} {\bibfnamefont {T.}~\bibnamefont
  {Mendes}},\ }\href {\doibase 10.1103/PhysRevD.78.094503} {\bibfield
  {journal} {\bibinfo  {journal} {Phys. Rev.}\ }\textbf {\bibinfo {volume}
  {D78}},\ \bibinfo {pages} {094503} (\bibinfo {year} {2008}{\natexlab{b}})},\
  \Eprint {http://arxiv.org/abs/0804.2371} {arXiv:0804.2371 [hep-lat]}
  \BibitemShut {NoStop}%
\bibitem [{\citenamefont {Dudal}\ \emph
  {et~al.}(2008{\natexlab{a}})\citenamefont {Dudal}, \citenamefont {Sorella},
  \citenamefont {Vandersickel},\ and\ \citenamefont
  {Verschelde}}]{Dudal:2007cw}%
  \BibitemOpen
  \bibfield  {author} {\bibinfo {author} {\bibfnamefont {D.}~\bibnamefont
  {Dudal}}, \bibinfo {author} {\bibfnamefont {S.~P.}\ \bibnamefont {Sorella}},
  \bibinfo {author} {\bibfnamefont {N.}~\bibnamefont {Vandersickel}}, \ and\
  \bibinfo {author} {\bibfnamefont {H.}~\bibnamefont {Verschelde}},\ }\href
  {\doibase 10.1103/PhysRevD.77.071501} {\bibfield  {journal} {\bibinfo
  {journal} {Phys. Rev.}\ }\textbf {\bibinfo {volume} {D77}},\ \bibinfo {pages}
  {071501} (\bibinfo {year} {2008}{\natexlab{a}})},\ \Eprint
  {http://arxiv.org/abs/0711.4496} {arXiv:0711.4496 [hep-th]} \BibitemShut
  {NoStop}%
\bibitem [{\citenamefont {Dudal}\ \emph
  {et~al.}(2008{\natexlab{b}})\citenamefont {Dudal}, \citenamefont {Gracey},
  \citenamefont {Sorella}, \citenamefont {Vandersickel},\ and\ \citenamefont
  {Verschelde}}]{Dudal:2008sp}%
  \BibitemOpen
  \bibfield  {author} {\bibinfo {author} {\bibfnamefont {D.}~\bibnamefont
  {Dudal}}, \bibinfo {author} {\bibfnamefont {J.~A.}\ \bibnamefont {Gracey}},
  \bibinfo {author} {\bibfnamefont {S.~P.}\ \bibnamefont {Sorella}}, \bibinfo
  {author} {\bibfnamefont {N.}~\bibnamefont {Vandersickel}}, \ and\ \bibinfo
  {author} {\bibfnamefont {H.}~\bibnamefont {Verschelde}},\ }\href {\doibase
  10.1103/PhysRevD.78.065047} {\bibfield  {journal} {\bibinfo  {journal} {Phys.
  Rev.}\ }\textbf {\bibinfo {volume} {D78}},\ \bibinfo {pages} {065047}
  (\bibinfo {year} {2008}{\natexlab{b}})},\ \Eprint
  {http://arxiv.org/abs/0806.4348} {arXiv:0806.4348 [hep-th]} \BibitemShut
  {NoStop}%
\bibitem [{\citenamefont {Aguilar}\ \emph {et~al.}(2008)\citenamefont
  {Aguilar}, \citenamefont {Binosi},\ and\ \citenamefont
  {Papavassiliou}}]{Aguilar:2008xm}%
  \BibitemOpen
  \bibfield  {author} {\bibinfo {author} {\bibfnamefont {A.~C.}\ \bibnamefont
  {Aguilar}}, \bibinfo {author} {\bibfnamefont {D.}~\bibnamefont {Binosi}}, \
  and\ \bibinfo {author} {\bibfnamefont {J.}~\bibnamefont {Papavassiliou}},\
  }\href {\doibase 10.1103/PhysRevD.78.025010} {\bibfield  {journal} {\bibinfo
  {journal} {Phys. Rev.}\ }\textbf {\bibinfo {volume} {D78}},\ \bibinfo {pages}
  {025010} (\bibinfo {year} {2008})},\ \Eprint {http://arxiv.org/abs/0802.1870}
  {arXiv:0802.1870 [hep-ph]} \BibitemShut {NoStop}%
\bibitem [{\citenamefont {Boucaud}\ \emph {et~al.}(2008)\citenamefont
  {Boucaud}, \citenamefont {Leroy}, \citenamefont {Yaouanc}, \citenamefont
  {Micheli}, \citenamefont {Pene},\ and\ \citenamefont
  {Rodriguez-Quintero}}]{Boucaud:2008ji}%
  \BibitemOpen
  \bibfield  {author} {\bibinfo {author} {\bibfnamefont {P.}~\bibnamefont
  {Boucaud}}, \bibinfo {author} {\bibfnamefont {J.-P.}\ \bibnamefont {Leroy}},
  \bibinfo {author} {\bibfnamefont {A.~L.}\ \bibnamefont {Yaouanc}}, \bibinfo
  {author} {\bibfnamefont {J.}~\bibnamefont {Micheli}}, \bibinfo {author}
  {\bibfnamefont {O.}~\bibnamefont {Pene}}, \ and\ \bibinfo {author}
  {\bibfnamefont {J.}~\bibnamefont {Rodriguez-Quintero}},\ }\href {\doibase
  10.1088/1126-6708/2008/06/012} {\bibfield  {journal} {\bibinfo  {journal}
  {JHEP}\ }\textbf {\bibinfo {volume} {06}},\ \bibinfo {pages} {012} (\bibinfo
  {year} {2008})},\ \Eprint {http://arxiv.org/abs/0801.2721} {arXiv:0801.2721
  [hep-ph]} \BibitemShut {NoStop}%
\bibitem [{\citenamefont {Cyrol}\ \emph {et~al.}(2016)\citenamefont {Cyrol},
  \citenamefont {Fister}, \citenamefont {Mitter}, \citenamefont {Pawlowski},\
  and\ \citenamefont {Strodthoff}}]{Cyrol:2016tym}%
  \BibitemOpen
  \bibfield  {author} {\bibinfo {author} {\bibfnamefont {A.~K.}\ \bibnamefont
  {Cyrol}}, \bibinfo {author} {\bibfnamefont {L.}~\bibnamefont {Fister}},
  \bibinfo {author} {\bibfnamefont {M.}~\bibnamefont {Mitter}}, \bibinfo
  {author} {\bibfnamefont {J.~M.}\ \bibnamefont {Pawlowski}}, \ and\ \bibinfo
  {author} {\bibfnamefont {N.}~\bibnamefont {Strodthoff}},\ }\href {\doibase
  10.1103/PhysRevD.94.054005} {\bibfield  {journal} {\bibinfo  {journal} {Phys.
  Rev.}\ }\textbf {\bibinfo {volume} {D94}},\ \bibinfo {pages} {054005}
  (\bibinfo {year} {2016})},\ \Eprint {http://arxiv.org/abs/1605.01856}
  {arXiv:1605.01856 [hep-ph]} \BibitemShut {NoStop}%
\bibitem [{\citenamefont {Dudal}\ \emph
  {et~al.}(2011{\natexlab{a}})\citenamefont {Dudal}, \citenamefont {Sorella},\
  and\ \citenamefont {Vandersickel}}]{Dudal:2011gd}%
  \BibitemOpen
  \bibfield  {author} {\bibinfo {author} {\bibfnamefont {D.}~\bibnamefont
  {Dudal}}, \bibinfo {author} {\bibfnamefont {S.~P.}\ \bibnamefont {Sorella}},
  \ and\ \bibinfo {author} {\bibfnamefont {N.}~\bibnamefont {Vandersickel}},\
  }\href {\doibase 10.1103/PhysRevD.84.065039} {\bibfield  {journal} {\bibinfo
  {journal} {Phys. Rev.}\ }\textbf {\bibinfo {volume} {D84}},\ \bibinfo {pages}
  {065039} (\bibinfo {year} {2011}{\natexlab{a}})},\ \Eprint
  {http://arxiv.org/abs/1105.3371} {arXiv:1105.3371 [hep-th]} \BibitemShut
  {NoStop}%
\bibitem [{\citenamefont {Capri}\ \emph {et~al.}(2015)\citenamefont {Capri},
  \citenamefont {Dudal}, \citenamefont {Fiorentini}, \citenamefont {Guimaraes},
  \citenamefont {Justo}, \citenamefont {Pereira}, \citenamefont {Mintz},
  \citenamefont {Palhares}, \citenamefont {Sobreiro},\ and\ \citenamefont
  {Sorella}}]{Capri:2015ixa}%
  \BibitemOpen
  \bibfield  {author} {\bibinfo {author} {\bibfnamefont {M.~A.~L.}\
  \bibnamefont {Capri}}, \bibinfo {author} {\bibfnamefont {D.}~\bibnamefont
  {Dudal}}, \bibinfo {author} {\bibfnamefont {D.}~\bibnamefont {Fiorentini}},
  \bibinfo {author} {\bibfnamefont {M.~S.}\ \bibnamefont {Guimaraes}}, \bibinfo
  {author} {\bibfnamefont {I.~F.}\ \bibnamefont {Justo}}, \bibinfo {author}
  {\bibfnamefont {A.~D.}\ \bibnamefont {Pereira}}, \bibinfo {author}
  {\bibfnamefont {B.~W.}\ \bibnamefont {Mintz}}, \bibinfo {author}
  {\bibfnamefont {L.~F.}\ \bibnamefont {Palhares}}, \bibinfo {author}
  {\bibfnamefont {R.~F.}\ \bibnamefont {Sobreiro}}, \ and\ \bibinfo {author}
  {\bibfnamefont {S.~P.}\ \bibnamefont {Sorella}},\ }\href {\doibase
  10.1103/PhysRevD.92.045039} {\bibfield  {journal} {\bibinfo  {journal} {Phys.
  Rev.}\ }\textbf {\bibinfo {volume} {D92}},\ \bibinfo {pages} {045039}
  (\bibinfo {year} {2015})},\ \Eprint {http://arxiv.org/abs/1506.06995}
  {arXiv:1506.06995 [hep-th]} \BibitemShut {NoStop}%
\bibitem [{\citenamefont {Capri}\ \emph
  {et~al.}(2016{\natexlab{a}})\citenamefont {Capri}, \citenamefont {Dudal},
  \citenamefont {Fiorentini}, \citenamefont {Guimaraes}, \citenamefont {Justo},
  \citenamefont {Pereira}, \citenamefont {Mintz}, \citenamefont {Palhares},
  \citenamefont {Sobreiro},\ and\ \citenamefont {Sorella}}]{Capri:2016aqq}%
  \BibitemOpen
  \bibfield  {author} {\bibinfo {author} {\bibfnamefont {M.~A.~L.}\
  \bibnamefont {Capri}}, \bibinfo {author} {\bibfnamefont {D.}~\bibnamefont
  {Dudal}}, \bibinfo {author} {\bibfnamefont {D.}~\bibnamefont {Fiorentini}},
  \bibinfo {author} {\bibfnamefont {M.~S.}\ \bibnamefont {Guimaraes}}, \bibinfo
  {author} {\bibfnamefont {I.~F.}\ \bibnamefont {Justo}}, \bibinfo {author}
  {\bibfnamefont {A.~D.}\ \bibnamefont {Pereira}}, \bibinfo {author}
  {\bibfnamefont {B.~W.}\ \bibnamefont {Mintz}}, \bibinfo {author}
  {\bibfnamefont {L.~F.}\ \bibnamefont {Palhares}}, \bibinfo {author}
  {\bibfnamefont {R.~F.}\ \bibnamefont {Sobreiro}}, \ and\ \bibinfo {author}
  {\bibfnamefont {S.~P.}\ \bibnamefont {Sorella}},\ }\href {\doibase
  10.1103/PhysRevD.94.025035} {\bibfield  {journal} {\bibinfo  {journal} {Phys.
  Rev.}\ }\textbf {\bibinfo {volume} {D94}},\ \bibinfo {pages} {025035}
  (\bibinfo {year} {2016}{\natexlab{a}})},\ \Eprint
  {http://arxiv.org/abs/1605.02610} {arXiv:1605.02610 [hep-th]} \BibitemShut
  {NoStop}%
\bibitem [{\citenamefont {Capri}\ \emph
  {et~al.}(2017{\natexlab{a}})\citenamefont {Capri}, \citenamefont {Dudal},
  \citenamefont {Pereira}, \citenamefont {Fiorentini}, \citenamefont
  {Guimaraes}, \citenamefont {Mintz}, \citenamefont {Palhares},\ and\
  \citenamefont {Sorella}}]{Capri:2016gut}%
  \BibitemOpen
  \bibfield  {author} {\bibinfo {author} {\bibfnamefont {M.~A.~L.}\
  \bibnamefont {Capri}}, \bibinfo {author} {\bibfnamefont {D.}~\bibnamefont
  {Dudal}}, \bibinfo {author} {\bibfnamefont {A.~D.}\ \bibnamefont {Pereira}},
  \bibinfo {author} {\bibfnamefont {D.}~\bibnamefont {Fiorentini}}, \bibinfo
  {author} {\bibfnamefont {M.~S.}\ \bibnamefont {Guimaraes}}, \bibinfo {author}
  {\bibfnamefont {B.~W.}\ \bibnamefont {Mintz}}, \bibinfo {author}
  {\bibfnamefont {L.~F.}\ \bibnamefont {Palhares}}, \ and\ \bibinfo {author}
  {\bibfnamefont {S.~P.}\ \bibnamefont {Sorella}},\ }\href {\doibase
  10.1103/PhysRevD.95.045011} {\bibfield  {journal} {\bibinfo  {journal} {Phys.
  Rev.}\ }\textbf {\bibinfo {volume} {D95}},\ \bibinfo {pages} {045011}
  (\bibinfo {year} {2017}{\natexlab{a}})},\ \Eprint
  {http://arxiv.org/abs/1611.10077} {arXiv:1611.10077 [hep-th]} \BibitemShut
  {NoStop}%
\bibitem [{\citenamefont {Capri}\ \emph
  {et~al.}(2017{\natexlab{b}})\citenamefont {Capri}, \citenamefont
  {Fiorentini}, \citenamefont {Pereira},\ and\ \citenamefont
  {Sorella}}]{Capri:2017abz}%
  \BibitemOpen
  \bibfield  {author} {\bibinfo {author} {\bibfnamefont {M.~A.~L.}\
  \bibnamefont {Capri}}, \bibinfo {author} {\bibfnamefont {D.}~\bibnamefont
  {Fiorentini}}, \bibinfo {author} {\bibfnamefont {A.~D.}\ \bibnamefont
  {Pereira}}, \ and\ \bibinfo {author} {\bibfnamefont {S.~P.}\ \bibnamefont
  {Sorella}},\ }\href {\doibase 10.1140/epjc/s10052-017-5107-z} {\bibfield
  {journal} {\bibinfo  {journal} {Eur. Phys. J.}\ }\textbf {\bibinfo {volume}
  {C77}},\ \bibinfo {pages} {546} (\bibinfo {year} {2017}{\natexlab{b}})},\
  \Eprint {http://arxiv.org/abs/1703.03264} {arXiv:1703.03264 [hep-th]}
  \BibitemShut {NoStop}%
\bibitem [{\citenamefont {Capri}\ \emph {et~al.}(2018)\citenamefont {Capri},
  \citenamefont {Dudal}, \citenamefont {Guimaraes}, \citenamefont {Pereira},
  \citenamefont {Mintz}, \citenamefont {Palhares},\ and\ \citenamefont
  {Sorella}}]{Capri:2018ijg}%
  \BibitemOpen
  \bibfield  {author} {\bibinfo {author} {\bibfnamefont {M.~A.~L.}\
  \bibnamefont {Capri}}, \bibinfo {author} {\bibfnamefont {D.}~\bibnamefont
  {Dudal}}, \bibinfo {author} {\bibfnamefont {M.~S.}\ \bibnamefont
  {Guimaraes}}, \bibinfo {author} {\bibfnamefont {A.~D.}\ \bibnamefont
  {Pereira}}, \bibinfo {author} {\bibfnamefont {B.~W.}\ \bibnamefont {Mintz}},
  \bibinfo {author} {\bibfnamefont {L.~F.}\ \bibnamefont {Palhares}}, \ and\
  \bibinfo {author} {\bibfnamefont {S.~P.}\ \bibnamefont {Sorella}},\ }\href
  {\doibase 10.1016/j.physletb.2018.03.058} {\bibfield  {journal} {\bibinfo
  {journal} {Phys. Lett.}\ }\textbf {\bibinfo {volume} {B781}},\ \bibinfo
  {pages} {48} (\bibinfo {year} {2018})},\ \Eprint
  {http://arxiv.org/abs/1802.04582} {arXiv:1802.04582 [hep-th]} \BibitemShut
  {NoStop}%
\bibitem [{\citenamefont {Delbourgo}\ and\ \citenamefont
  {Thompson}(1986)}]{Delbourgo:1986wz}%
  \BibitemOpen
  \bibfield  {author} {\bibinfo {author} {\bibfnamefont {R.}~\bibnamefont
  {Delbourgo}}\ and\ \bibinfo {author} {\bibfnamefont {G.}~\bibnamefont
  {Thompson}},\ }\href {\doibase 10.1103/PhysRevLett.57.2610} {\bibfield
  {journal} {\bibinfo  {journal} {Phys. Rev. Lett.}\ }\textbf {\bibinfo
  {volume} {57}},\ \bibinfo {pages} {2610} (\bibinfo {year}
  {1986})}\BibitemShut {NoStop}%
\bibitem [{\citenamefont {Lavelle}\ and\ \citenamefont
  {McMullan}(1997)}]{Lavelle:1995ty}%
  \BibitemOpen
  \bibfield  {author} {\bibinfo {author} {\bibfnamefont {M.}~\bibnamefont
  {Lavelle}}\ and\ \bibinfo {author} {\bibfnamefont {D.}~\bibnamefont
  {McMullan}},\ }\href {\doibase 10.1016/S0370-1573(96)00019-1} {\bibfield
  {journal} {\bibinfo  {journal} {Phys. Rept.}\ }\textbf {\bibinfo {volume}
  {279}},\ \bibinfo {pages} {1} (\bibinfo {year} {1997})},\ \Eprint
  {http://arxiv.org/abs/hep-ph/9509344} {arXiv:hep-ph/9509344 [hep-ph]}
  \BibitemShut {NoStop}%
\bibitem [{\citenamefont {Capri}\ \emph
  {et~al.}(2016{\natexlab{b}})\citenamefont {Capri}, \citenamefont
  {Fiorentini}, \citenamefont {Guimaraes}, \citenamefont {Mintz}, \citenamefont
  {Palhares}, \citenamefont {Sorella}, \citenamefont {Dudal}, \citenamefont
  {Justo}, \citenamefont {Pereira},\ and\ \citenamefont
  {Sobreiro}}]{Capri:2015nzw}%
  \BibitemOpen
  \bibfield  {author} {\bibinfo {author} {\bibfnamefont {M.~A.~L.}\
  \bibnamefont {Capri}}, \bibinfo {author} {\bibfnamefont {D.}~\bibnamefont
  {Fiorentini}}, \bibinfo {author} {\bibfnamefont {M.~S.}\ \bibnamefont
  {Guimaraes}}, \bibinfo {author} {\bibfnamefont {B.~W.}\ \bibnamefont
  {Mintz}}, \bibinfo {author} {\bibfnamefont {L.~F.}\ \bibnamefont {Palhares}},
  \bibinfo {author} {\bibfnamefont {S.~P.}\ \bibnamefont {Sorella}}, \bibinfo
  {author} {\bibfnamefont {D.}~\bibnamefont {Dudal}}, \bibinfo {author}
  {\bibfnamefont {I.~F.}\ \bibnamefont {Justo}}, \bibinfo {author}
  {\bibfnamefont {A.~D.}\ \bibnamefont {Pereira}}, \ and\ \bibinfo {author}
  {\bibfnamefont {R.~F.}\ \bibnamefont {Sobreiro}},\ }\href {\doibase
  10.1103/PhysRevD.93.065019} {\bibfield  {journal} {\bibinfo  {journal} {Phys.
  Rev.}\ }\textbf {\bibinfo {volume} {D93}},\ \bibinfo {pages} {065019}
  (\bibinfo {year} {2016}{\natexlab{b}})},\ \Eprint
  {http://arxiv.org/abs/1512.05833} {arXiv:1512.05833 [hep-th]} \BibitemShut
  {NoStop}%
\bibitem [{\citenamefont {Capri}\ \emph
  {et~al.}(2016{\natexlab{c}})\citenamefont {Capri}, \citenamefont
  {Fiorentini}, \citenamefont {Guimaraes}, \citenamefont {Mintz}, \citenamefont
  {Palhares},\ and\ \citenamefont {Sorella}}]{Fiorentini:2016rwx}%
  \BibitemOpen
  \bibfield  {author} {\bibinfo {author} {\bibfnamefont {M.~A.~L.}\
  \bibnamefont {Capri}}, \bibinfo {author} {\bibfnamefont {D.}~\bibnamefont
  {Fiorentini}}, \bibinfo {author} {\bibfnamefont {M.~S.}\ \bibnamefont
  {Guimaraes}}, \bibinfo {author} {\bibfnamefont {B.~W.}\ \bibnamefont
  {Mintz}}, \bibinfo {author} {\bibfnamefont {L.~F.}\ \bibnamefont {Palhares}},
  \ and\ \bibinfo {author} {\bibfnamefont {S.~P.}\ \bibnamefont {Sorella}},\
  }\href {\doibase 10.1103/PhysRevD.94.065009} {\bibfield  {journal} {\bibinfo
  {journal} {Phys. Rev.}\ }\textbf {\bibinfo {volume} {D94}},\ \bibinfo {pages}
  {065009} (\bibinfo {year} {2016}{\natexlab{c}})},\ \Eprint
  {http://arxiv.org/abs/1606.06601} {arXiv:1606.06601 [hep-th]} \BibitemShut
  {NoStop}%
\bibitem [{\citenamefont {Capri}\ \emph
  {et~al.}(2017{\natexlab{c}})\citenamefont {Capri}, \citenamefont
  {Fiorentini}, \citenamefont {Pereira},\ and\ \citenamefont
  {Sorella}}]{Capri:2017bfd}%
  \BibitemOpen
  \bibfield  {author} {\bibinfo {author} {\bibfnamefont {M.~A.~L.}\
  \bibnamefont {Capri}}, \bibinfo {author} {\bibfnamefont {D.}~\bibnamefont
  {Fiorentini}}, \bibinfo {author} {\bibfnamefont {A.~D.}\ \bibnamefont
  {Pereira}}, \ and\ \bibinfo {author} {\bibfnamefont {S.~P.}\ \bibnamefont
  {Sorella}},\ }\href {\doibase 10.1103/PhysRevD.96.054022} {\bibfield
  {journal} {\bibinfo  {journal} {Phys. Rev.}\ }\textbf {\bibinfo {volume}
  {D96}},\ \bibinfo {pages} {054022} (\bibinfo {year} {2017}{\natexlab{c}})},\
  \Eprint {http://arxiv.org/abs/1708.01543} {arXiv:1708.01543 [hep-th]}
  \BibitemShut {NoStop}%
\bibitem [{Note2()}]{Note2}%
  \BibitemOpen
  \bibinfo {note} {With an important difference that the operators which
  condense in our case are not elementary fields that enter the lagrangian, as
  would be the case for spontaneous magnetization in a spin model.}\BibitemShut
  {Stop}%
\bibitem [{\citenamefont {Verschelde}\ \emph {et~al.}(2001)\citenamefont
  {Verschelde}, \citenamefont {Knecht}, \citenamefont {Van~Acoleyen},\ and\
  \citenamefont {Vanderkelen}}]{Verschelde:2001ia}%
  \BibitemOpen
  \bibfield  {author} {\bibinfo {author} {\bibfnamefont {H.}~\bibnamefont
  {Verschelde}}, \bibinfo {author} {\bibfnamefont {K.}~\bibnamefont {Knecht}},
  \bibinfo {author} {\bibfnamefont {K.}~\bibnamefont {Van~Acoleyen}}, \ and\
  \bibinfo {author} {\bibfnamefont {M.}~\bibnamefont {Vanderkelen}},\ }\href
  {\doibase 10.1016/S0370-2693(01)00929-7} {\bibfield  {journal} {\bibinfo
  {journal} {Phys. Lett.}\ }\textbf {\bibinfo {volume} {B516}},\ \bibinfo
  {pages} {307} (\bibinfo {year} {2001})},\ \Eprint
  {http://arxiv.org/abs/hep-th/0105018} {arXiv:hep-th/0105018 [hep-th]}
  \BibitemShut {NoStop}%
\bibitem [{\citenamefont {Knecht}\ and\ \citenamefont
  {Verschelde}(2001)}]{Knecht:2001cc}%
  \BibitemOpen
  \bibfield  {author} {\bibinfo {author} {\bibfnamefont {K.}~\bibnamefont
  {Knecht}}\ and\ \bibinfo {author} {\bibfnamefont {H.}~\bibnamefont
  {Verschelde}},\ }\href {\doibase 10.1103/PhysRevD.64.085006} {\bibfield
  {journal} {\bibinfo  {journal} {Phys. Rev.}\ }\textbf {\bibinfo {volume}
  {D64}},\ \bibinfo {pages} {085006} (\bibinfo {year} {2001})},\ \Eprint
  {http://arxiv.org/abs/hep-th/0104007} {arXiv:hep-th/0104007 [hep-th]}
  \BibitemShut {NoStop}%
\bibitem [{\citenamefont {Dudal}\ \emph {et~al.}(2003)\citenamefont {Dudal},
  \citenamefont {Verschelde}, \citenamefont {Browne},\ and\ \citenamefont
  {Gracey}}]{Dudal:2003vv}%
  \BibitemOpen
  \bibfield  {author} {\bibinfo {author} {\bibfnamefont {D.}~\bibnamefont
  {Dudal}}, \bibinfo {author} {\bibfnamefont {H.}~\bibnamefont {Verschelde}},
  \bibinfo {author} {\bibfnamefont {R.~E.}\ \bibnamefont {Browne}}, \ and\
  \bibinfo {author} {\bibfnamefont {J.~A.}\ \bibnamefont {Gracey}},\ }\href
  {\doibase 10.1016/S0370-2693(03)00541-0} {\bibfield  {journal} {\bibinfo
  {journal} {Phys. Lett.}\ }\textbf {\bibinfo {volume} {B562}},\ \bibinfo
  {pages} {87} (\bibinfo {year} {2003})},\ \Eprint
  {http://arxiv.org/abs/hep-th/0302128} {arXiv:hep-th/0302128 [hep-th]}
  \BibitemShut {NoStop}%
\bibitem [{\citenamefont {Browne}\ and\ \citenamefont
  {Gracey}(2003)}]{Browne:2003uv}%
  \BibitemOpen
  \bibfield  {author} {\bibinfo {author} {\bibfnamefont {R.~E.}\ \bibnamefont
  {Browne}}\ and\ \bibinfo {author} {\bibfnamefont {J.~A.}\ \bibnamefont
  {Gracey}},\ }\href {\doibase 10.1088/1126-6708/2003/11/029} {\bibfield
  {journal} {\bibinfo  {journal} {JHEP}\ }\textbf {\bibinfo {volume} {11}},\
  \bibinfo {pages} {029} (\bibinfo {year} {2003})},\ \Eprint
  {http://arxiv.org/abs/hep-th/0306200} {arXiv:hep-th/0306200 [hep-th]}
  \BibitemShut {NoStop}%
\bibitem [{\citenamefont {Dudal}\ \emph {et~al.}(2005)\citenamefont {Dudal},
  \citenamefont {Sobreiro}, \citenamefont {Sorella},\ and\ \citenamefont
  {Verschelde}}]{Dudal:2005na}%
  \BibitemOpen
  \bibfield  {author} {\bibinfo {author} {\bibfnamefont {D.}~\bibnamefont
  {Dudal}}, \bibinfo {author} {\bibfnamefont {R.~F.}\ \bibnamefont {Sobreiro}},
  \bibinfo {author} {\bibfnamefont {S.~P.}\ \bibnamefont {Sorella}}, \ and\
  \bibinfo {author} {\bibfnamefont {H.}~\bibnamefont {Verschelde}},\ }\href
  {\doibase 10.1103/PhysRevD.72.014016} {\bibfield  {journal} {\bibinfo
  {journal} {Phys. Rev.}\ }\textbf {\bibinfo {volume} {D72}},\ \bibinfo {pages}
  {014016} (\bibinfo {year} {2005})},\ \Eprint
  {http://arxiv.org/abs/hep-th/0502183} {arXiv:hep-th/0502183 [hep-th]}
  \BibitemShut {NoStop}%
\bibitem [{\citenamefont {Dudal}\ \emph {et~al.}(2018)\citenamefont {Dudal},
  \citenamefont {Felix}, \citenamefont {Palhares}, \citenamefont {Rondeau},\
  and\ \citenamefont {Vercauteren}}]{Dudal:2018ctc}%
  \BibitemOpen
  \bibfield  {author} {\bibinfo {author} {\bibfnamefont {D.}~\bibnamefont
  {Dudal}}, \bibinfo {author} {\bibfnamefont {C.}~\bibnamefont {Felix}},
  \bibinfo {author} {\bibfnamefont {L.}~\bibnamefont {Palhares}}, \bibinfo
  {author} {\bibfnamefont {F.}~\bibnamefont {Rondeau}}, \ and\ \bibinfo
  {author} {\bibfnamefont {D.}~\bibnamefont {Vercauteren}},\ }\href@noop {} {\
  (\bibinfo {year} {2018})},\ \Eprint {http://arxiv.org/abs/1811.12524}
  {arXiv:1811.12524 [hep-ph]} \BibitemShut {NoStop}%
\bibitem [{\citenamefont {Mermin}\ and\ \citenamefont
  {Wagner}(1966)}]{MerminWagner}%
  \BibitemOpen
  \bibfield  {author} {\bibinfo {author} {\bibfnamefont {N.~D.}\ \bibnamefont
  {Mermin}}\ and\ \bibinfo {author} {\bibfnamefont {H.}~\bibnamefont
  {Wagner}},\ }\href {\doibase 10.1103/PhysRevLett.17.1133} {\bibfield
  {journal} {\bibinfo  {journal} {Phys. Rev. Lett.}\ }\textbf {\bibinfo
  {volume} {17}},\ \bibinfo {pages} {1133} (\bibinfo {year}
  {1966})}\BibitemShut {NoStop}%
\bibitem [{\citenamefont {Coleman}(1973)}]{Coleman:1973ci}%
  \BibitemOpen
  \bibfield  {author} {\bibinfo {author} {\bibfnamefont {S.~R.}\ \bibnamefont
  {Coleman}},\ }\href {\doibase 10.1007/BF01646487} {\bibfield  {journal}
  {\bibinfo  {journal} {Commun. Math. Phys.}\ }\textbf {\bibinfo {volume}
  {31}},\ \bibinfo {pages} {259} (\bibinfo {year} {1973})}\BibitemShut
  {NoStop}%
\bibitem [{\citenamefont {Dudal}\ \emph {et~al.}(2014)\citenamefont {Dudal},
  \citenamefont {Guimaraes},\ and\ \citenamefont {Sorella}}]{Dudal:2013wja}%
  \BibitemOpen
  \bibfield  {author} {\bibinfo {author} {\bibfnamefont {D.}~\bibnamefont
  {Dudal}}, \bibinfo {author} {\bibfnamefont {M.~S.}\ \bibnamefont
  {Guimaraes}}, \ and\ \bibinfo {author} {\bibfnamefont {S.~P.}\ \bibnamefont
  {Sorella}},\ }\href {\doibase 10.1016/j.physletb.2014.03.056} {\bibfield
  {journal} {\bibinfo  {journal} {Phys. Lett.}\ }\textbf {\bibinfo {volume}
  {B732}},\ \bibinfo {pages} {247} (\bibinfo {year} {2014})},\ \Eprint
  {http://arxiv.org/abs/1310.2016} {arXiv:1310.2016 [hep-ph]} \BibitemShut
  {NoStop}%
\bibitem [{\citenamefont {Dudal}\ \emph
  {et~al.}(2011{\natexlab{b}})\citenamefont {Dudal}, \citenamefont
  {Guimaraes},\ and\ \citenamefont {Sorella}}]{Dudal:2010cd}%
  \BibitemOpen
  \bibfield  {author} {\bibinfo {author} {\bibfnamefont {D.}~\bibnamefont
  {Dudal}}, \bibinfo {author} {\bibfnamefont {M.~S.}\ \bibnamefont
  {Guimaraes}}, \ and\ \bibinfo {author} {\bibfnamefont {S.~P.}\ \bibnamefont
  {Sorella}},\ }\href {\doibase 10.1103/PhysRevLett.106.062003} {\bibfield
  {journal} {\bibinfo  {journal} {Phys. Rev. Lett.}\ }\textbf {\bibinfo
  {volume} {106}},\ \bibinfo {pages} {062003} (\bibinfo {year}
  {2011}{\natexlab{b}})},\ \Eprint {http://arxiv.org/abs/1010.3638}
  {arXiv:1010.3638 [hep-th]} \BibitemShut {NoStop}%
\bibitem [{\citenamefont {Pene}\ \emph {et~al.}(2010)\citenamefont {Pene} \emph
  {et~al.}}]{Pene:2011kg}%
  \BibitemOpen
  \bibfield  {author} {\bibinfo {author} {\bibfnamefont {O.}~\bibnamefont
  {Pene}} \emph {et~al.},\ }\bibfield  {booktitle} {\emph {\bibinfo {booktitle}
  {{Proceedings, Workshop on The many faces of QCD (FacesQCD2010): Ghent,
  Belgium, November 1-5, 2010}}},\ }\href {\doibase 10.22323/1.117.0010}
  {\bibfield  {journal} {\bibinfo  {journal} {PoS}\ }\textbf {\bibinfo {volume}
  {FACESQCD}},\ \bibinfo {pages} {010} (\bibinfo {year} {2010})},\ \Eprint
  {http://arxiv.org/abs/1102.1535} {arXiv:1102.1535 [hep-lat]} \BibitemShut
  {NoStop}%
\bibitem [{\citenamefont {Chernodub}\ and\ \citenamefont
  {Ilgenfritz}(2008)}]{Chernodub:2008kf}%
  \BibitemOpen
  \bibfield  {author} {\bibinfo {author} {\bibfnamefont {M.~N.}\ \bibnamefont
  {Chernodub}}\ and\ \bibinfo {author} {\bibfnamefont {E.~M.}\ \bibnamefont
  {Ilgenfritz}},\ }\href {\doibase 10.1103/PhysRevD.78.034036} {\bibfield
  {journal} {\bibinfo  {journal} {Phys. Rev.}\ }\textbf {\bibinfo {volume}
  {D78}},\ \bibinfo {pages} {034036} (\bibinfo {year} {2008})},\ \Eprint
  {http://arxiv.org/abs/0805.3714} {arXiv:0805.3714 [hep-lat]} \BibitemShut
  {NoStop}%
\end{thebibliography}%

\end{document}